\documentclass[symmetry,article,accept,moreauthors,pdftex]{Definitions/mdpi}

\firstpage{1}
\pubvolume{12}
\issuenum{4}
\articlenumber{648}
\pubyear{2020}
\copyrightyear{2020}
\history{Received: 30 March 2020; Accepted: 15 April 2020; Published: 20 April 2020}
\updates{yes} 

\usepackage[utf8]{inputenc}
\usepackage[T1]{fontenc}
\usepackage{amssymb}
\usepackage{mathtools}
\usepackage{subcaption}
\usepackage{textcomp,gensymb}
\usepackage{listings}
\newcommand{\arcdeg}{\(\stackrel{\:\circ}{\textstyle.\rule{0pt}{0.40ex}}\)}

\Title{The Axisymmetric Central Configurations of the Four-Body Problem with Three Equal Masses}

\Author{Emese K{\H o}v\'ari *\href{https://orcid.org/0000-0002-4491-2824}{\orcidicon} and B{\'a}lint {\'E}rdi}

\AuthorNames{Emese K{\H o}v\'ari and B{\'a}lint {\'E}rdi}

\address[1]{%
Department of Astronomy, Eötvös Loránd University, Pázmány Péter sétány 1/A,  1117 Budapest, Hungary; b.erdi@astro.elte.hu}

\corres{\hspace{-.75EM}Correspondence: e.kovari@astro.elte.hu}

\abstract{\textls[-5]{In the studied axisymmetric case of the central four-body problem, the axis of symmetry is defined by two unequal-mass bodies, while the other two bodies are situated symmetrically with respect to this axis and have equal masses. Here, we consider a special case of the problem and assume that three of the masses are equal. Using a recently found analytical solution of the general case, we formulate the equations of condition for three equal masses analytically and solve them numerically. A complete description of the problem is given by providing both the coordinates and masses of the bodies. We show furthermore how the three-equal-mass solutions are related to the general case in the coordinate space. The physical aspects of the configurations are also studied and~discussed.}}


\begin{document}

\section{Introduction}
\label{introduction}
A fundamental problem in celestial mechanics is to describe the motion of $ n $ pointlike bodies, assuming only the mutual Newtonian gravitational forces acting between them. Given a set of masses $ m_i $ and barycentric position vectors $ {\boldsymbol r}_i $, the equations of motion take the form
\begin{equation}
\label{eq:nbodyprob}
m_i \ddot{\boldsymbol r}_i =  \displaystyle\sum_{\substack{j = 1\\ j \neq i}}^n\frac{m_i m_j}{r_{ij}^3}\left ({\boldsymbol r}_j - {\boldsymbol r}_i\right ), \ \ \ \ \ \ \ i = 1, \dots, n,
\end{equation}
where $ r_{ij} \equiv |\boldsymbol{r}_i - \boldsymbol{r}_j| $ is the distance between the $ i $-th and $ j $-th bodies. It is assumed that the units of mass, length, and time are such that the constant of gravity is $ 1 $.
\par
System \eqref{eq:nbodyprob} is generally unsolvable for more than two bodies, which gives rise to the study of an important subclass, i.e., central configurations. A configuration is called central if the resulting forces on each body are directed toward the center of mass of the system, or in a mathematical formulation, if there exists some $ \lambda \in {\rm I\!R} $, $ \lambda > 0 $ such that $ \ddot{\boldsymbol r}_i = -\lambda {\boldsymbol r}_i $ for all $ i = 1, \dots , n $. Accordingly, in the central case, Equation \eqref{eq:nbodyprob} can be rewritten as
\begin{equation}
\label{eq:centr}
\displaystyle\sum_{\substack{j = 1\\ j \neq i}}^n\frac{m_j}{r_{ij}^3}\left ({\boldsymbol r}_j - {\boldsymbol r}_i\right ) = -\lambda {\boldsymbol r}_i, \ \ \ \ \ \ \ i = 1, \dots, n.
\end{equation}

The importance of central configurations can be approached from several different perspectives. From a mathematical point of view, central configurations represent the only subclass of the $ n $-body problem that provides explicit analytic solutions for all time. Moreover, these solutions have the interesting property of being self-similar, i.e., invariant under rotation, translation, and dilation. Central configurations also arise in regard to the determination of the topology of the $ n $-body problem (see \cite{meyer1987,smale1970} for further details). From a physical viewpoint, one must mention both total-collisions and total gravitational expansions (all the bodies of the system collide or expand at a given time), whose configurations also tend to central ones \cite{saari1980,diacu1992,dziobek1990}.

\textls[-5]{Notwithstanding the above instances, the study of central configurations is still greatly challenging. For small values of $ n $, the solutions of Equation \eqref{eq:centr} were already given in the 17th and 18th centuries (see the Newtonian two-body problem for $ n = 2 $, and the collinear and equilateral triangle solutions for $ n = 3 $, found by Euler and Lagrange). For larger values of $ n $ however, we face a notoriously complicated problem, highlighted by the fact that for $ n \ge 4 $ the number of the different types of central configurations---and for $ n \ge 6 $ even the finiteness of this number---is still a pending~question.}

In the $ n = 4 $ case, a full characterization of the so-called kite central configurations was given recently by Érdi and Czirják \cite{erdi2016}. They studied the case where an axis of symmetry connects two bodies and the other two bodies of equal masses are situated symmetrically with respect to this axis---the four bodies thus forming a (convex or concave) deltoid (the geometric form of a kite).

\textls[-8]{The problem of central configurations of four bodies with three equal masses has been studied in several papers. It is known from the literature that the concave kite configurations of three equal masses must be either an equilateral triangle formed by the equal-mass bodies and the fourth mass occupying its geometric center, or an isosceles triangle with the fourth body on the axis of symmetry of the triangle~\cite{palmore1975,simo1978,meyer1988,long2002,bernat2009}. In the case of convex configurations, it was proved \cite{long2002} that if three masses are equal, then the bodies must necessarily form a kite configuration.  In the planar symmetric case of the four-body problem with three equal masses, it was found \cite{shi2010} that besides the equilateral triangle configuration, there exists one group of concave and one group of convex central configurations. It~is worth noting that in the concave case, there exist nonkite central configurations with three equal masses as well \cite{bernat2009}.}\enlargethispage{0.5cm}

\textls[-5]{In this paper, we aim to show how the analytical solutions given in \cite{erdi2016} can be applied to systems with three equal masses. According to these solutions, the kite central configurations can be parametrized by two angles, and such configurations can only exist for angles in some well-determined domains of the parameter plane of the angles. The two masses outside the axis of symmetry are always equal; therefore, the cases of three equal masses occur when either one (or both) of the masses on the axis of symmetry become(s) equal to them. Using the analytical formulae given in \cite{erdi2016}, we first derive the analytical conditions of these cases; then, by numerical computations, represent them as solution curves in the allowed domains of the parameter plane, thus giving a clear view of how the three-equal-mass cases are related to the general case of the axisymmetric central configurations of four bodies. This latter relation was previously not known. We also compute numerically the masses of the bodies and discuss how they change along the solution curves in the allowed domains. We compare and discuss our solutions with respect to previous results.}

The paper is organized as follows. In Section~\ref{sec:conditions_for_the_three_equal_masses}, we derive the equations of condition for the cases of three equal masses. The numerical solutions of these equations are presented in Section~\ref{sec:solutions_in_the_beta_alpha_plane}, describing the families of solutions for the convex and concave cases separately. We compute the masses in Section~\ref{sec:computing_the_non_dimensional_masses} and discuss their changes along the solution curves. A summary of the results is given in Section~\ref{sec:summary}. Some of the analytical considerations are detailed in the Appendixes \ref{sec:appendix_a}--\ref{sec:appendix_c}.

\section{Equations of Condition for Three Equal Masses}
\label{sec:conditions_for_the_three_equal_masses}
\begin{figure}[H]
\centering
\includegraphics[width = 0.47\textwidth]{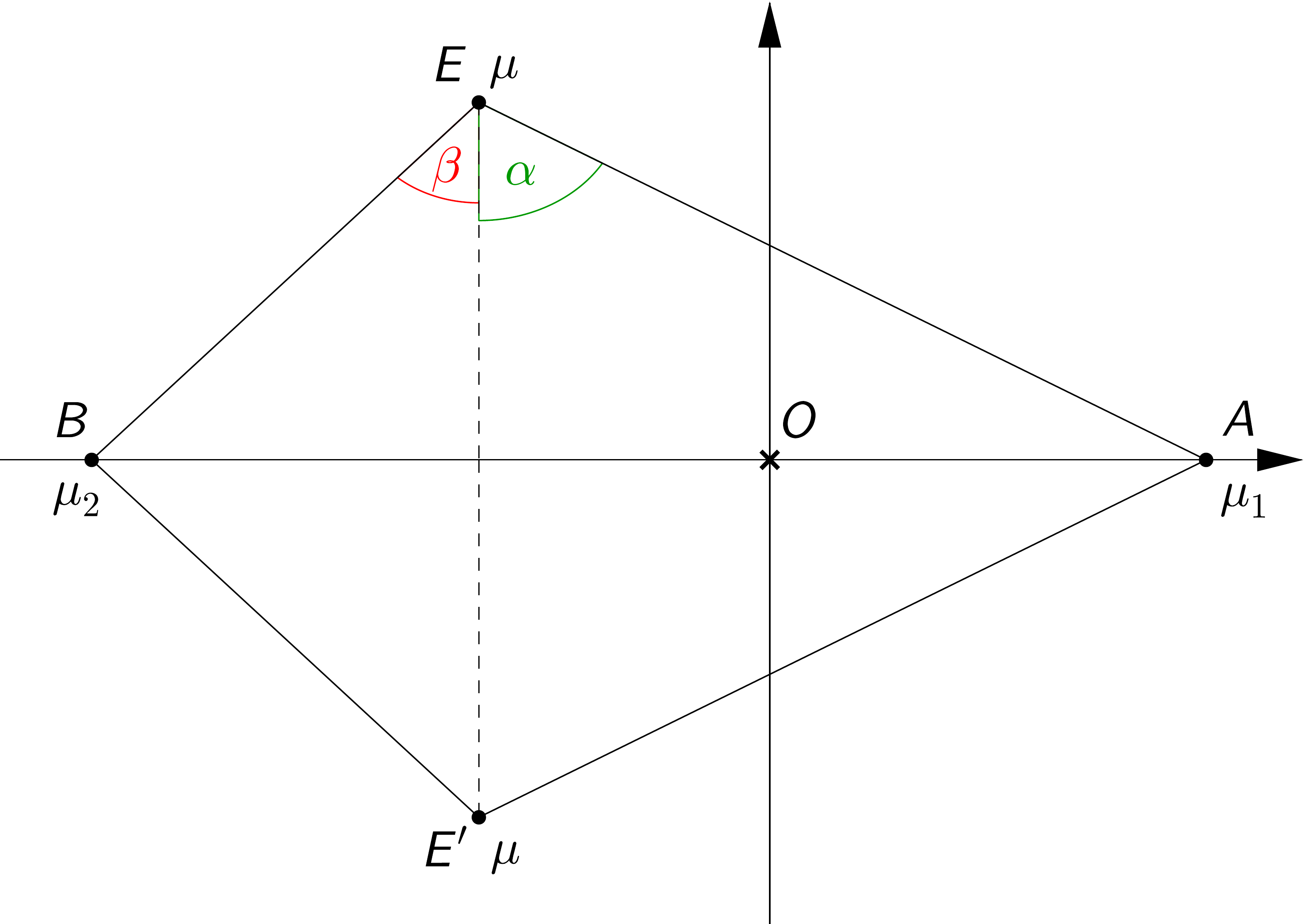}
\caption{Convex central configuration. The four bodies, forming a convex deltoid, are labeled by $ A $, $ B $, $ E $, and $ E' $, and their nondimensional masses are $ \mu_1 $, $ \mu_2 $, $ \mu $, and $ \mu $, respectively. $ O $ indicates the center of mass of the system. The two independent parameters are the angle coordinates $ \alpha $ and $ \beta $.}
\label{fig:cx_def}
\end{figure}

\begin{figure}[H]
\centering
\begin{subfigure}[!h]{0.47\textwidth}
\includegraphics[width=\textwidth]{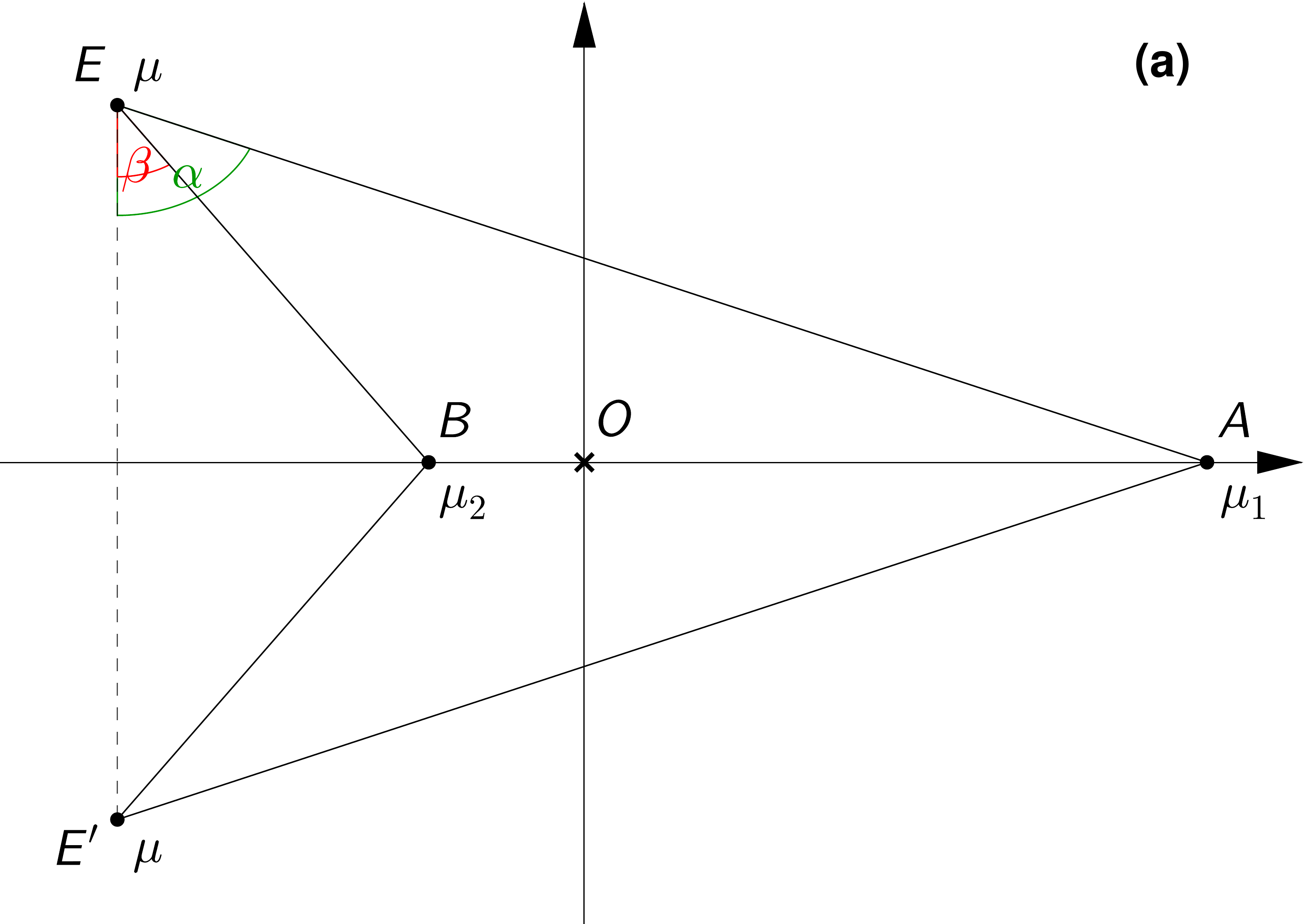}
\end{subfigure}
\quad
\begin{subfigure}[!h]{0.47\textwidth}
\includegraphics[width=\textwidth]{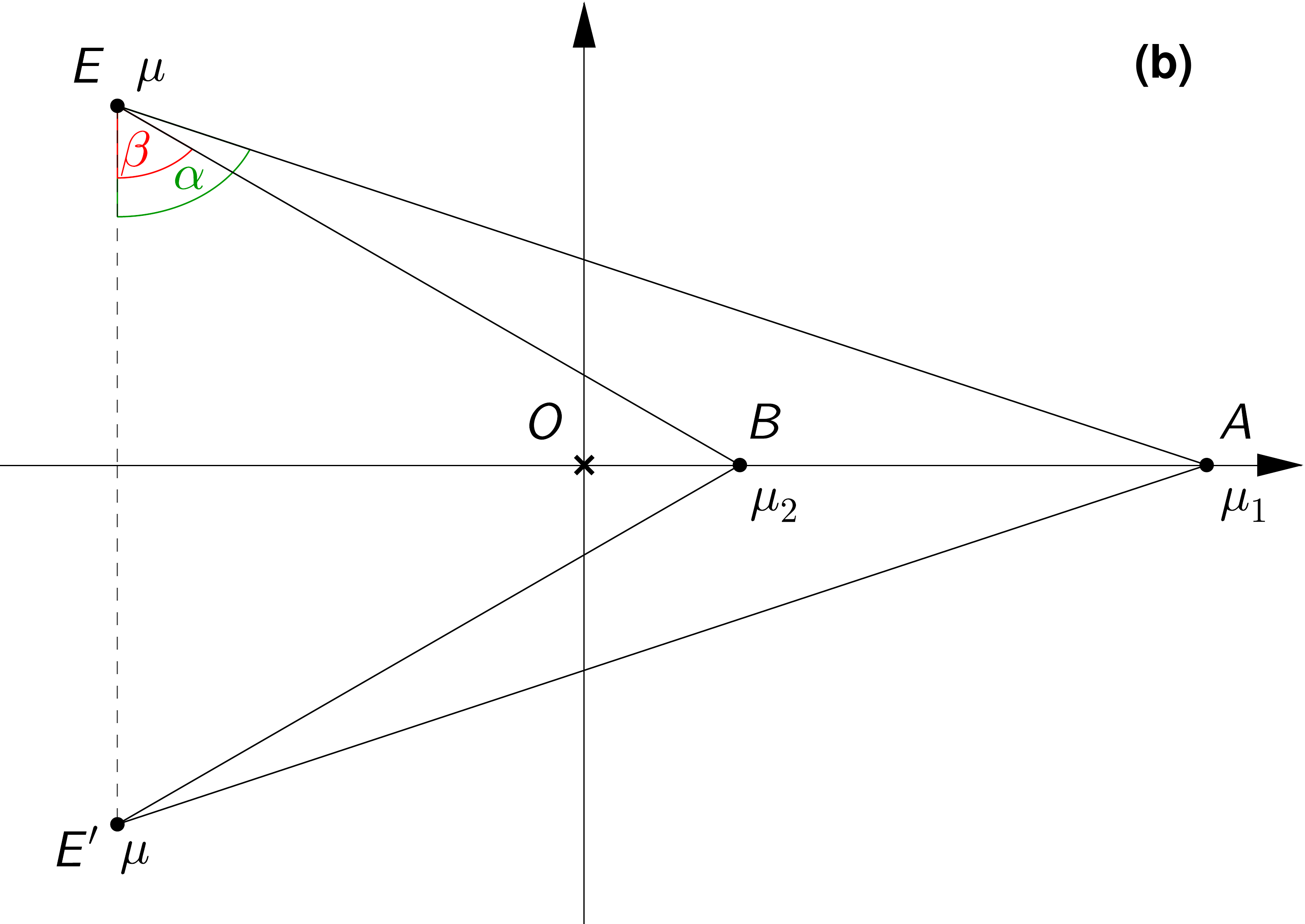}
\end{subfigure}
\caption{Concave central configurations. The denotations are the same as in {Figure} \ref{fig:cx_def}. The two independent parameters are the angle coordinates $ \alpha $ and $ \beta $. (\textbf{a}) First concave case. The center of mass $ O $ is inside the deltoid. (\textbf{b}) Second concave case. $ O $ is outside of the deltoid.}
\label{fig:cv_def}
\end{figure}

Using the same frame as in \cite{erdi2016}, we consider four mass-points, denoted by $ A $, $ B $, $ E $, and $ E' $ (see Figures \ref{fig:cx_def} and \ref{fig:cv_def}). It is assumed that the axis of symmetry of the system connects the bodies $ A $ and $ B $, whereas $ E $ and $ E' $ of equal masses are situated symmetrically with respect to this axis. The~nondimensional masses (see Equations (36) and (37) of \cite{erdi2016}) are designated in a way that $ A $ has mass $ \mu_1 $, $ B $ has mass $ \mu_2 $, and both $ E $ and $ E' $ have mass $ \mu $. (The mass $ \mu $ can be obtained as $ \mu = (1-\mu_1-\mu_2)/2 $.) {Figure} \ref{fig:cx_def} shows the convex case, where the four mass-points form a convex deltoid. Due to reasons of symmetry, it can be assumed that $\mu_2 \leq \mu_1$. {Figure} \ref{fig:cv_def} shows the two concave cases, distinguished by the criterion that in the first case ({Figure} \ref{fig:cv_def}a), the center of mass $O$ of the system is inside the concave deltoid formed by the four mass-points; while in the second case ({Figure} \ref{fig:cv_def}b), it is outside of it.
\unskip

Equation (88) in \cite{erdi2016} gives the masses in the form
\begin{equation}
\label{eq:Solmu1mu2}
\mu_1=\frac{(b_1+a_0-b_0)b_0}{a_0b_1+a_1b_0-a_1b_1}, \qquad
\mu_2=\frac{(a_1+b_0-a_0)a_0}{a_0b_1+a_1b_0-a_1b_1},
\end{equation}
where the coefficients $a_0$, $a_1$, $b_0$, $b_1$ are trigonometric expressions of the angles $\alpha$ and $\beta$, defined as shown in Figures \ref{fig:cx_def} and \ref{fig:cv_def} (see Equations (54)--(57) for the convex cases and Equations (66)--(69) for the concave cases
of \cite{erdi2016}). For convenience, they are repeated here in Appendix \ref{sec:appendix_a}. We note that using Equation \eqref{eq:Solmu1mu2}, the masses should satisfy the conditions $ 0 \leq \mu_1 \leq 1 $, $ 0 \leq \mu_2 \leq 1 $, and $ 0 \leq \mu_1 + \mu_2 \leq 1 $. We also note that in the concave cases, there is a constraint on the angle $ \beta $, requiring that $ \tan \beta $ is  either smaller (first case), or larger (second case) than $ (\mu_1\tan \alpha) /(1-\mu_2) $.

Searching for configurations with three equal masses and considering that both $ E $ and $ E' $ have mass $ \mu $, the desired conditions are $ \mu = \mu_1 $ or $ \mu = \mu_2 $. Next, we consider these cases. (When $ \mu = \mu_1 = \mu_2 $, all four masses are equal. These are special cases of the three-equal-mass solutions.)

\subsection{Case $ \mu = \mu_1 $}
\label{subsec:case_mu_equals_mu1}
Considering that $ \mu = (1-\mu_1-\mu_2)/2 $, the condition $ \mu = \mu_1 $ yields $ 3 \mu_1 + \mu_2 = 1 $. Using the solutions \eqref{eq:Solmu1mu2}, we obtain that
\begin{equation}
\label{eq:muequalsmu1}
3 \frac{(b_1 + a_0 - b_0)b_0}{a _0 b_1 + a_1 b_0 - a_1 b_1}
+ \frac{(a_1 + b_0 - a_0)a_0}{a_0 b_1 + a_1 b_0 - a_1 b_1} = 1.
\end{equation}

Multiplying by the denominators, then simplifying with the term $ b_1 +a_0 - b_0 \neq 0 $ (see Section~\ref{subsubsec:excep1}), Equation \eqref{eq:muequalsmu1} leads to
\begin{equation}
\label{eq:muequalsmu1final}
a_0 - a_1 - 3b_0 = 0.
\end{equation}

Substituting here the coefficients from Appendix \ref{sec:appendix_a}, after some arrangements, we obtain in the convex case that
\begin{equation}
\label{eq:muequalsmu1convexfinal}
\tan \beta \left (\cos^3 \alpha - 2\cos^3 \beta + \frac{1}{4} \right )
+ \tan \alpha \cos^3 \alpha - \frac{1}{(\tan \alpha + \tan \beta)^2} = 0,
\end{equation}
and in the concave cases that
\begin{equation}
\label{eq:muequalsmu1concavefinal}
- \tan \beta \left (\cos^3 \alpha - 2\cos^3 \beta + \frac{1}{4} \right )
+ \tan \alpha \cos^3 \alpha - \frac{1}{(\tan \alpha - \tan \beta)^2} = 0.
\end{equation}

It is worth noting that Equations \eqref{eq:muequalsmu1convexfinal} and \eqref{eq:muequalsmu1concavefinal} differ only in the sign of $ \tan \beta $. This property is related to the different introductions of the angle $ \beta $ in the convex and concave cases (see Figures \ref{fig:cx_def} and \ref{fig:cv_def}).

\subsection{Case $ \mu = \mu_2 $}
\label{subsec:case_mu_equals_mu2}
The equations in this case can be derived similarly as in the previous section. We start from the criterion $ 3 \mu_2 + \mu_1 = 1 $, which takes the form
\begin{equation}
\label{eq:muequalsmu2}
3 \frac{(a_1 + b_0 - a_0)a_0}{a_0 b_1 + a_1 b_0 - a_1 b_1}
+ \frac{(b_1 + a_0 - b_0)b_0}{a_0 b_1 + a_1 b_0 - a_1 b_1} = 1,
\end{equation}
after using the solutions \eqref{eq:Solmu1mu2} for the masses. Simplifying with the term $ a_1 + b_0 -a_0 \neq 0 $ (see Section~\ref{subsubsec:excep2}), it follows that
\begin{equation}
\label{eq:muequalsmu2final}
b_0 - b_1 - 3a_0 = 0.
\end{equation}

Substituting again the coefficients from Appendix \ref{sec:appendix_a}, after some arrangements, we obtain in the convex case that
\begin{equation}
\label{eq:muequalsmu2convexfinal}
\tan \alpha \left (\cos^3 \beta - 2\cos^3 \alpha + \frac{1}{4} \right )
+ \tan \beta \cos^3 \beta - \frac{1}{(\tan \alpha + \tan \beta)^2} = 0,
\end{equation}
and in the concave cases that
\begin{equation}
\label{eq:muequalsmu2concavefinal2}
\tan \alpha \left (\cos^3 \beta - 2\cos^3 \alpha + \frac{1}{4} \right )
- \tan \beta \cos^3 \beta - \frac{1}{(\tan \alpha - \tan \beta)^2} = 0.
\end{equation}
\par
The only difference between Equations \eqref{eq:muequalsmu2convexfinal} and \eqref{eq:muequalsmu2concavefinal2} is the sign of $ \tan \beta $, similarly to Equations \eqref{eq:muequalsmu1convexfinal} and \eqref{eq:muequalsmu1concavefinal}.

\subsection{Exceptional Cases}
\label{subsec:exceptional_cases}
We assumed previously that both $ b_1 + a_0 - b_0 $ and  $ a_1 + b_0 - a_0 $ are nonzero terms. Now, we investigate what happens if we equate them to $ 0 $.

\subsubsection{$ b_1 +a_0 - b_0 = 0 $}
\label{subsubsec:excep1}
With this assumption, Equations \eqref{eq:Solmu1mu2} and \eqref{eq:muequalsmu1} lead to $ \mu = \mu_1 = 0 $, and consequently $ \mu_2 = 1 $.

This is not allowed in the convex case, since by assumption, $ \mu_2 \leq \mu_1 $.

In the concave cases however, there exist such solutions, namely, the points of the line $ 2 \alpha - \beta = 90 \degree $, along which $ \mu_1 = 0 $, $ \mu_2 = 1 $ (for more detail, see Section 7.2.1 of \cite{erdi2016}).~This satisfies the condition $ 3 \mu_1 + \mu_2 = 1 $.\enlargethispage{0.5cm}

\subsubsection{$ a_1 + b_0 -a_0 = 0 $}
\label{subsubsec:excep2}
In this case, Equations \eqref{eq:Solmu1mu2} and \eqref{eq:muequalsmu2} lead to $ \mu = \mu_2 = 0 $, and consequently, $ \mu_1 = 1 $.

In the convex case, there exist such solutions, namely, the points of the line $ \alpha + 2 \beta = 90 \degree $, along which $ \mu_1 = 1 $, $ \mu_2 = 0 $ (for more detail, 
see Section 6.1.1 of \cite{erdi2016}). This satisfies the condition $ 3 \mu_2 + \mu_1 = 1 $.

In the concave cases, the above condition leads to the equation  $ \cos \beta = \sin (\alpha - \beta) $, which is satisfied if and only if $ \beta $ and $ \alpha - \beta $ are complementary angles, that is, $ \beta + \alpha - \beta = \alpha = 90 \degree $. However, this case is not allowed, since $\alpha < 90\degree$ according to its definition (see {Figure} \ref{fig:cv_def}).

\section{Solutions in the $ \beta $, $ \alpha $ Plane}
\label{sec:solutions_in_the_beta_alpha_plane}
With the lack of analytical solutions of Equations \eqref{eq:muequalsmu1convexfinal}, \eqref{eq:muequalsmu1concavefinal}, \eqref{eq:muequalsmu2convexfinal}, and \eqref{eq:muequalsmu2concavefinal2}, we used \emph{MATLAB} to determine the points on the $\beta$, $\alpha$ parameter plane that satisfy these equations. The regions where the kite central configurations of four bodies can exist are already known from \cite{erdi2016} and are also shown here as the colored areas in {Figure} \ref{fig:betaalphaplane_cx} (for the convex) and in {Figure} \ref{fig:betaalphaplane_cv} (for the two concave cases). Therefore, we only considered solutions in the given domains. The resulting curves appear in the $ \beta $, $ \alpha $ parameter plane as the graphs, i.e., the $ \left(\beta, \alpha(\beta)\right) $ pairs, of such implicit functions $ \beta \mapsto f_i(\beta) \equiv \alpha(\beta) $ $(i = 1, \dots, 4) $ that satisfy the above equations. In the following, we describe our results separately for the convex and concave cases.

\subsection{Convex Case}
\label{subsec:solutions_in_the_beta_alpha_plane_cx}
\begin{figure}[H]
\centering
\includegraphics[width = 0.65\textwidth]{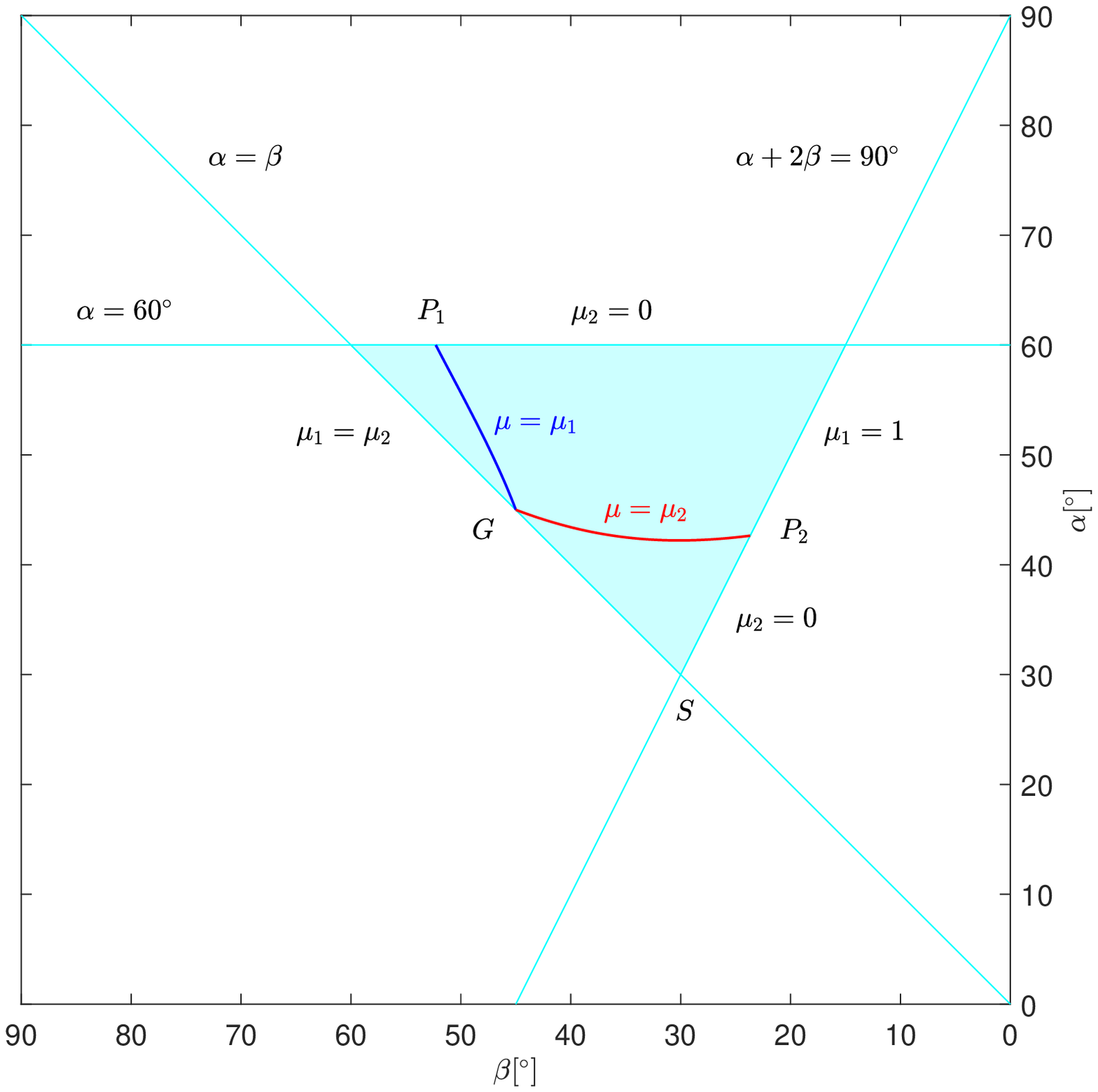}
\caption{Solutions in the convex case. The colored area (bordered by critical lines) refers to those $ \beta $, $ \alpha $ pairs that allow convex central configurations of four bodies. Letter $S$ at $\alpha=\beta=30 \degree$ indicates a singular point where $\mu_1+\mu_2=1$. Letter $ G $ at $ \alpha = \beta = 45 \degree $ corresponds to the square configuration with four equal masses. Along the blue curve, the bodies $ A $, $ E $, and $ E' $ have equal masses, while the masses of $ B $, $ E $, and $ E' $ become equal along the red curve. Letters $ P_1 $ and $ P_2 $ indicate the endpoints of the solutions curves on critical lines. Note the reversed direction of the horizontal axis.}
\label{fig:betaalphaplane_cx}
\end{figure}

The solutions in the convex case are displayed in {Figure} \ref{fig:betaalphaplane_cx}. All convex central configurations, with the assumptions of \cite{erdi2016}, are restricted to the interior of a triangle (colored area in {Figure} \ref{fig:betaalphaplane_cx}), bordered by the critical lines $ \alpha = 60 \degree $, $ \alpha + 2 \beta = 90 \degree $, and $ \alpha = \beta$, along which the masses take extremal values as indicated in the figure.

The square configuration \cite{albouy1996} with four equal masses at $ \alpha = \beta = 45 \degree $ (point $ G $ in {Figure} \ref{fig:betaalphaplane_cx}) can be considered as a generating configuration of the three-equal-mass cases, since both the $ \mu = \mu_1 $ and $ \mu = \mu_2 $ solutions start from here. (The solution points $ \alpha = \beta = 45 \degree $ are obtained either from Equation~\eqref{eq:muequalsmu1convexfinal} or from Equation \eqref{eq:muequalsmu2convexfinal} by substituting $ \alpha = \beta $.)

In the $ \mu = \mu_1 $ case (blue curve in {Figure} \ref{fig:betaalphaplane_cx}), the masses of the bodies $ A $, $ E $, and $ E' $ are equal, and as $ \alpha $ increases from $ 45 \degree $ to $ 60 \degree $ and $ \beta $ from $ 45 \degree $ to 52\arcdeg282, the initial square configuration is replaced by a convex deltoid, which is more and more elongated along its line of symmetry. The endpoint $ P_1 $ in {Figure} \ref{fig:betaalphaplane_cx} is on the critical line $ \alpha = 60 \degree $, and the value of $ \beta $ for this point is obtained from Equation \eqref{eq:muequalsmu1convexfinal} by substituting $ \alpha = 60 \degree $. One observes that $ \mu_2 $ is zero along this critical line, thus, at the endpoint $ P_1 $ of the blue curve, the four-body problem is reduced to a three-body problem with the bodies $ A $, $ E $, and $ E' $ forming an equilateral triangle and having the common mass $1/3$. We note here an important property, namely that in the convex $ \mu = \mu_1 $ case, for each $ \alpha $ there exists a unique $ \beta $ satisfying the condition of three equal masses. In other words, it means that the function $ f_1 $, associated with the blue curve, is injective, for which it is enough to be strictly monotone. The proof of this is given in Appendix \ref{subsec:appendix_b1}.

In the $ \mu = \mu_2 $ case, the solution is represented by a red curve in {Figure} \ref{fig:betaalphaplane_cx}. In contrast to the previous case, the corresponding function $ f_2 $ has a local minimum at $ \alpha = \text{42\arcdeg211} $, $ \beta = \text{30\arcdeg153} $ (see Appendix \ref{subsec:appendix_b2}), implying it to be noninjective. It follows from this finding that between this minimum point and the endpoint of the red curve on the critical line $ \alpha + 2 \beta = 90 \degree $ (point $ P_2 $ in {Figure} \ref{fig:betaalphaplane_cx}), for each value of $ \alpha $ there exist two values of $ \beta $ for which $ \mu=\mu_2 $. This interval of $ \alpha $ is rather narrow, but the corresponding interval of $ \beta $ is wide: $ \text{23\arcdeg680} < \beta < \text{36\arcdeg125} $. For example, for $ \alpha = \text{42\arcdeg500} $ there is a $ \beta = \text{24\arcdeg883} $ with corresponding masses $ \mu = \mu_2 = 0.0426 $, $ \mu_1 = 0.8723 $; and another $ \beta = \text{35\arcdeg081} $ with corresponding masses $ \mu = \mu_2 = 0.1914 $, $ \mu_1 = 0.4259 $. The values of $ \alpha = \text{42\arcdeg639} $ and $ \beta = \text{23\arcdeg680} $ at the endpoint $ P_2 $ can be computed from Equation \eqref{eq:muequalsmu2convexfinal} by substituting $ \alpha = 90 \degree - 2 \beta $ as obtained from the equation of the respective critical line. At this endpoint, the masses are $ \mu_1 = 1 $, $ \mu_2 = 0 $; consequently, $ B $, $ E $, and $ E' $ become zero-mass bodies, and the configuration is reduced to a one-body problem with $ A $ of mass $ 1 $.

\textls[-5]{In \cite{bernat2009}, the central configurations of four bodies with three equal masses were studied by a different method and from a different point of view. As a main result, the authors determined the number of all possible types of configurations depending on the ratio of the nonequal mass with respect to the equal mass of the bodies. Here, we study the problem of three equal masses with respect to the general case of the axisymmetric central configurations of four bodies, with an emphasis on the physical aspects of the configurations. Though the methods and the scopes are different in the two studies---in~\cite{bernat2009}, the problem was studied in itself---some of the results can be compared. The abovementioned points $ P_1 $ and $ P_2 $ appear in our approach as intersections of the solution curves and two of the critical lines in {Figure} \ref{fig:betaalphaplane_cx}. They are also among the 7 special points of \cite{bernat2009}: $ P_1 \equiv (k, l) = (\sqrt{3}, 1.29302) $ and $ P_2 = (0.43856, 0.92080) $. By the transformation $ \tan \alpha = k $, $ \tan \beta = l $ between the angles $\alpha$, $\beta$ and the coordinates $k$, $l$ of \cite{bernat2009}, the identity of the two solutions can be seen.}

\subsection{Concave Cases}
\label{subsec:solutions_in_the_beta_alpha_plane_cv}
\begin{figure}[H]
\centering
\includegraphics[width = 0.65\textwidth]{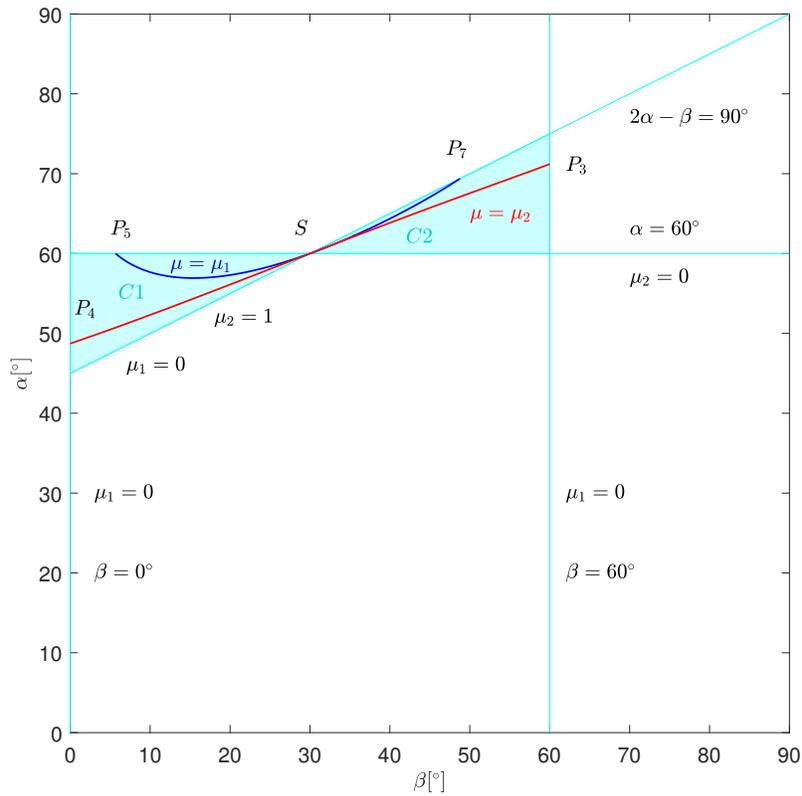}
\caption{Solutions in the concave cases. The colored areas (bordered by critical lines) refer to those $ \beta $, $ \alpha $ pairs that allow concave central configurations of four bodies. The triangles of the first and second concave cases are denoted by $ C1 $ and $ C2 $, respectively. Letter $ S $ at $ \alpha = 60 \degree $, $ \beta = 30 \degree $ indicates a singular point where $ 3 \mu_1 + \mu_2 = 1 $. Along the blue curve, the bodies $ A $, $ E $, and $ E' $ have equal masses, while the masses of $ B $, $ E $, and $ E' $ become equal along the red curve. The two curves have two intersections, one at the singular point $ S $, the other one at $ \alpha = \text{61\arcdeg177} $, $ \beta = \text{33\arcdeg039} $ (see a magnification in Figure \ref{fig:41_betaalphaplane_concaves_zoom}). Letters $ P_3 $, $ P_4 $, $ P_5 $, and $ P_7 $ indicate the endpoints of the solutions curves on critical lines.}
\label{fig:betaalphaplane_cv}
\end{figure}

The results in the concave cases are shown in {Figure} \ref{fig:betaalphaplane_cv}. According to \cite{erdi2016}, the concave kite central configurations  are confined to the interiors of two triangles (colored areas in {Figure} \ref{fig:betaalphaplane_cv}, $ C1 $ and $ C2 $ denoting the first and second concave cases, respectively). The bordering critical lines are $ \beta = 0 \degree $, $ \alpha = 60 \degree $, $ 2 \alpha - \beta = 90 \degree $, and $ \beta = 60 \degree $, where the masses take extremal values as indicated in the figure.

\begin{figure}[H]
\centering
\includegraphics[width=0.65\textwidth]{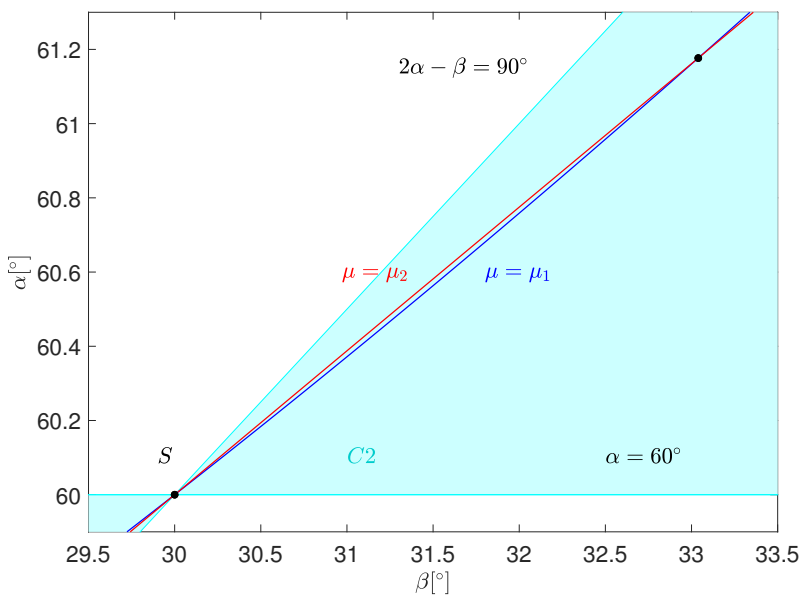}
\caption{Magnification of the intersections of the blue $ \mu = \mu_1 $ and  red $ \mu = \mu_2 $ curves, indicated by black dots. Both occur in the second concave case ($ C2 $); the first at $ \alpha = 60 \degree $, $ \beta = 30 \degree $ and the second at $ \alpha = \text{61\arcdeg177} $, $ \beta = \text{33\arcdeg039} $.}
\label{fig:41_betaalphaplane_concaves_zoom}
\end{figure}

The common point of the two triangles, labeled by $ S $, refers to a singular point, where $ \mu_1 $ and $ \mu_2 $ are individually undetermined but for the sum of them holds the condition $ 3 \mu_1 + \mu_2 = 1 $. From the definition of $ \mu $, it also follows that $ \mu = \mu_1 $ at the singular point, thus the masses of $ A $, $ E $, and $ E'$ are equal for any values of $ 0 \leq \mu_2 \leq 1 $, satisfying the above condition. Therefore, the blue curve, corresponding to the $\mu=\mu_1$ case, necessarily goes through $S$. However, the red curve that corresponds to the $\mu=\mu_2$ case also crosses this point, since $\mu=\mu_2$ is also possible here---that is, when all four masses are equal: $\mu=\mu_1=\mu_2= 1/4 $. In both cases, $A$, $E$, and $E'$ form an equilateral triangle (since $\alpha=60\degree$) and $ B $ occupies its center (since $\beta=30\degree$). We note that apart from the singular point $ S $, there exists one more point allowing four equal masses (see also \cite{albouy1996}), namely, at $ \alpha = \text{61\arcdeg177} $, $ \beta = \text{33\arcdeg039} $, where the blue and red curves cross each other. (The above values were obtained by equating the left-hand sides of Equations \eqref{eq:muequalsmu1concavefinal} and \eqref{eq:muequalsmu2concavefinal2}.) The two curves stay very close after point $S$; therefore, we show a magnification of the two intersections in {Figure} \ref{fig:41_betaalphaplane_concaves_zoom}.

\unskip
The conditions $\mu=\mu_1$ and $\mu=\mu_2$ have solutions both in the first and second concave cases.

The blue curve in {Figure} \ref{fig:betaalphaplane_cv} represents the solutions of $\mu=\mu_1$. Its starting point is labeled by $ P_5 $, and it is on the critical line $ \alpha = 60 \degree $ in the first concave region. The other coordinate $ \beta = \text{5\arcdeg678} $ was computed from Equation \eqref{eq:muequalsmu1concavefinal}, using the substitution $ \alpha = 60 \degree $. At this point, $ \mu_2 $ is zero (as is all along the critical line), and thus the common mass of the bodies $ A $, $ E $, and $ E' $ is necessarily  $ \mu = \mu_1 = 1/3 $. The~configuration is a Lagrangian equilateral triangle. After leaving $ P_5 $, the blue curve (associated with function $ f_3 $) first decreases to a minimum at $ \alpha = \text{56\arcdeg930}$, $ \beta =\text{15\arcdeg414} $, then it increases again. Since the initial value of $ \alpha = 60 \degree $ is reached again at the singular point $ S $, the property that for each value of $\alpha$ there exist two values of $\beta$ for which $\mu=\mu_1$, holds along the blue curve in the whole first concave region. For example, for $\alpha=58\degree$, there is a $\beta=\text{9\arcdeg059} $ with corresponding masses $\mu=\mu_1=0.3242$, $\mu_2=0.0273$; and another $\beta=\text{23\arcdeg460}$ with corresponding masses $\mu=\mu_1=0.2889$, $\mu_2=0.1332$. (This noninjectivity of $ f_3 $ in the first concave region is shown in Appendix \ref{subsec:appendix_b3}) In the second concave case, the blue curve continues to increase (the proof is given in Appendix \ref{subsec:appendix_b3} by showing the positive sign of the derivative of $ f_3 $ in this region), and it ends on the critical line $ 2 \alpha - \beta = 90 \degree $. The coordinates $ \alpha = \text{69\arcdeg383} $, $ \beta = \text{48\arcdeg765} $ of this endpoint ($ P_7 $ in {Figure} \ref{fig:betaalphaplane_cv}) can be computed from Equation \eqref{eq:muequalsmu1concavefinal} with the substitution $ \alpha = (90 \degree + \beta)/2 $. In this point, $ \mu_1 = 0 $ and $ \mu_2 = 1 $, in accordance with the property of the respective critical line, thus the masses of $ A $, $ E $, and  $ E' $ are 0 and the system is reduced to a one-body problem of $ B $ with unit mass.

The red curve in {Figure} \ref{fig:betaalphaplane_cv} represents the solutions of $\mu=\mu_2$. Its starting point is labeled by $ P_4 $ and it is on the critical line $ \beta = 0 \degree $. The other coordinate $ \alpha = \text{48\arcdeg729} $ was computed from Equation \eqref{eq:muequalsmu2concavefinal2} using the substitution $ \beta = 0 \degree $. At this point, $ \mu_1 $ is zero (as is all along the critical line), and thus the common mass of the bodies $ B $, $ E $, and $ E' $ is $ \mu = \mu_2=1/3 $. The configuration is a Eulerian collinear three-body problem. In contrast to the previous $ \mu = \mu_1 $ case, the function $ f_4 $---associated with the red curve---strictly monotonically increases in all points of its domain of definition, implying to be injective both in the first and second concave regions. (See Appendix \ref{subsec:appendix_b4} for the proof.) The endpoint ($ P_3 $ in {Figure} \ref{fig:betaalphaplane_cv}) falls on the critical line $ \beta = 60 \degree $, thus its other coordinate $ \alpha = \text{71\arcdeg199} $ is obtained from Equation \eqref{eq:muequalsmu2concavefinal2} with the above substitution. In this point, $ \mu_1 = 0 $, in accordance with the property of the respective critical line, thus the masses of $ B $, $ E $, and $ E' $ are $ 1/3 $ again, but now they form a Lagrangian equilateral triangle.

The endpoints $ P_3 $, $ P_4 $, $ P_5 $, and $ P_7$, as the intersections of the solution curves and critical lines in {Figure} \ref{fig:betaalphaplane_cv}, are related to the special points of \cite{bernat2009}: $ P_3 \equiv (k, l) = (-\sqrt{3}, 2.93734) $, $ P_4 = (0, 1.13942) $, $ P_5 = (\sqrt{3}, -0.09943) $, and $ P_7 = (2.65802, -1.14090) $. In the case of $ \mu = \mu_1 $ (concerning the points $ P_5 $ and $ P_7 $), the transformation between the angle coordinates $ \alpha $, $ \beta $ and the coordinates $ k $, $ l $ of \cite{bernat2009} is $ \tan \alpha = k $, $ \tan \beta = -l $. In the case of $ \mu = \mu_2 $ (regarding the points $ P_3 $ and $ P_4 $), the transformation is $ \tan \alpha = l $, $ \tan \beta = -k $. We note that the point $P_6$ of \cite{bernat2009} corresponds to the singular point $S$ in {Figure} \ref{fig:betaalphaplane_cv}.

\section{Computing the Nondimensional Masses}
\label{sec:computing_the_non_dimensional_masses}
In order to provide a complete description of the problem, we computed the nondimensional masses from Equation \eqref{eq:Solmu1mu2} for the angle coordinates obtained from Equations \eqref{eq:muequalsmu1convexfinal}, \eqref{eq:muequalsmu1concavefinal},  \eqref{eq:muequalsmu2convexfinal}, and \eqref{eq:muequalsmu2concavefinal2}. In~the following, we study the masses and their changes in the function of $ \beta $ separately for the convex and concave cases.

\subsection{Convex Case}
\label{subsec:computing_the_non_dimensional_masses_cx}
\begin{figure}[H]
\centering
\includegraphics[width = 0.48\textwidth]{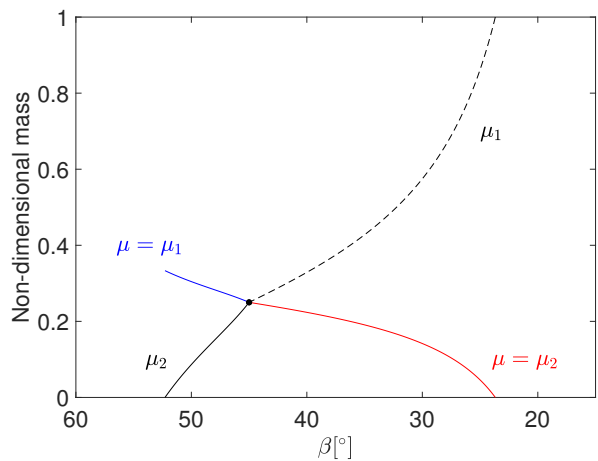}
\caption{Normalized masses in the function of the angle coordinate $ \beta $ (convex case). Both the $ \mu = \mu_1 $ (blue) and $ \mu = \mu_2 $ (red) solutions are displayed. The black solid and dashed curves refer to the residual mass $\mu_2$ or $\mu_1$ as opposed to the three equal masses. The generating point at $ \beta = 45 \degree $ is emphasized by a black dot and indicates the common mass $ \mu = \mu_1 = \mu_2 = 1/4 $. Note the reversed direction of the horizontal axis.}
\label{fig:beta_mass_cx}
\end{figure}

The a priori observation in Section~\ref{subsec:solutions_in_the_beta_alpha_plane_cx} that the intersection of the curves $ \mu = \mu_1 $ and $ \mu = \mu_2 $ at $ \alpha = \beta = 45 \degree $ is a generating solution with four equal masses is seen again in {Figure} \ref{fig:beta_mass_cx}. The common nondimensional mass is $ 1/4 $ at this point.

In the case of $ \mu = \mu_1 $ (blue curve), we leave this point with increasing values of $ \beta $ (and $ \alpha $) as the common mass of the bodies $ A $, $ E $, and $ E' $ approaches $1/3 $ and reaches it at $ \beta = \text{52\arcdeg282} $ ($ \alpha = 60 \degree $), forcing $ B $, after a gradual decline, to have zero mass.

In the case of $ \mu = \mu_2 $ (red curve), the common mass of the bodies $ B $, $ E $, and $ E' $ is decreasing from $ 1/4 $ to $0$ as $ \beta $ decreases from $ 45 \degree $ to $ \text{23\arcdeg680} $. This allows the mass of $ A $ to increase and reach $ 1 $ at the endpoint ($\beta = \text{23\arcdeg680} $, $\alpha= \text{42\arcdeg639} $).

\subsection{Concave Cases}
\label{subsec:computing_the_non_dimensional_masses_cv}
\begin{figure}[H]
\centering
\includegraphics[width = 0.48\textwidth]{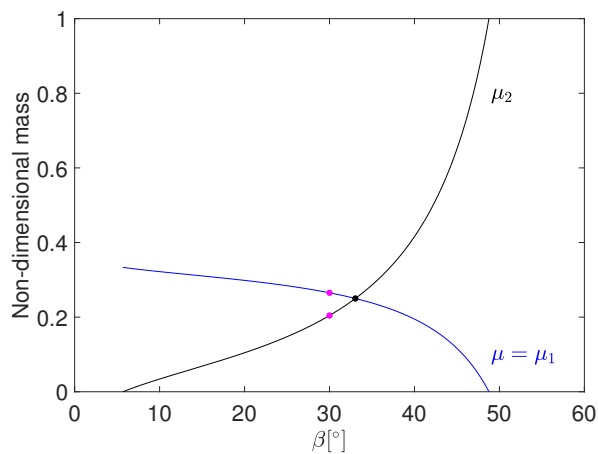}
\caption{Normalized masses in the function of the angle coordinate $ \beta $ (concave cases, $ \mu = \mu_1 $). The black solid line refers to the residual mass $ \mu_2 $. The pink dots at $ \beta = 30 \degree $ indicate the masses corresponding to Palmore's constant. The intersection at $ \beta = \text{33\arcdeg039} $ is emphasized by a black dot and indicates the common mass $ \mu = \mu_1 = \mu_2 = 1/4 $.}
\label{fig:beta_mass_cv1}
\end{figure}

In the $ \mu = \mu_1 $ case ({Figure} \ref{fig:beta_mass_cv1}, blue curve), the minimal value of $ \beta = \text{5\arcdeg678} $ occurs at $ \alpha = 60 \degree $ (see {Figure} \ref{fig:betaalphaplane_cv}), i.e., the critical masses are $ \mu = \mu_1 =1/3 $ and $ \mu_2= 0 $, whereas the configuration is a Lagrangian equilateral triangle of the bodies $ A $, $ E $, and $ E' $. As $ \beta $ increases, $ \mu_2 $ begins to increase too, reaching the final unit mass at $ \beta = \text{48\arcdeg765} $, while the masses $\mu=\mu_1$ undergo a decreasing tendency toward 0 at the endpoint. Along the solution curves, two points deserve special attention. At $\beta=30\degree$, corresponding to the singular point $S$, the values of the masses are $\mu=\mu_1=0.26522$, $\mu_2=0.20435$ (indicated by pink dots in {Figure} \ref{fig:beta_mass_cv1}), whose ratio $\mu_2/\mu_1=0.77049$ is Palmore's constant \mbox{$P=(2+3\sqrt{3})/(18-5\sqrt{3})$}. Palmore proved \cite{palmore1975} that the family of equilateral triangle central configurations of four bodies, with three equal masses at the vertices and an arbitrary fourth mass at the center of the triangle, is degenerate for the mass ratio $P$ of the central to the outer bodies. In \cite{meyer1988}, it was shown that an isosceles triangle family bifurcates from the degenerate equilateral configuration, with three equal-mass bodies at the vertices of the isosceles triangle, and the fourth body on the axis of symmetry of the triangle. {Figure} \ref{fig:beta_mass_cv1} shows the masses of this isosceles triangle family both in the first ($\beta<30\degree$) and second ($\beta>30\degree$) concave cases. The second remarkable point in {Figure} \ref{fig:beta_mass_cv1} is the intersection of the two curves at $\beta= \text{33\arcdeg039} $, corresponding to a solution of four equal masses \cite{albouy1996}.

\begin{figure}[H]
\centering
\begin{subfigure}[!h]{0.48\textwidth}
\includegraphics[width=\textwidth]{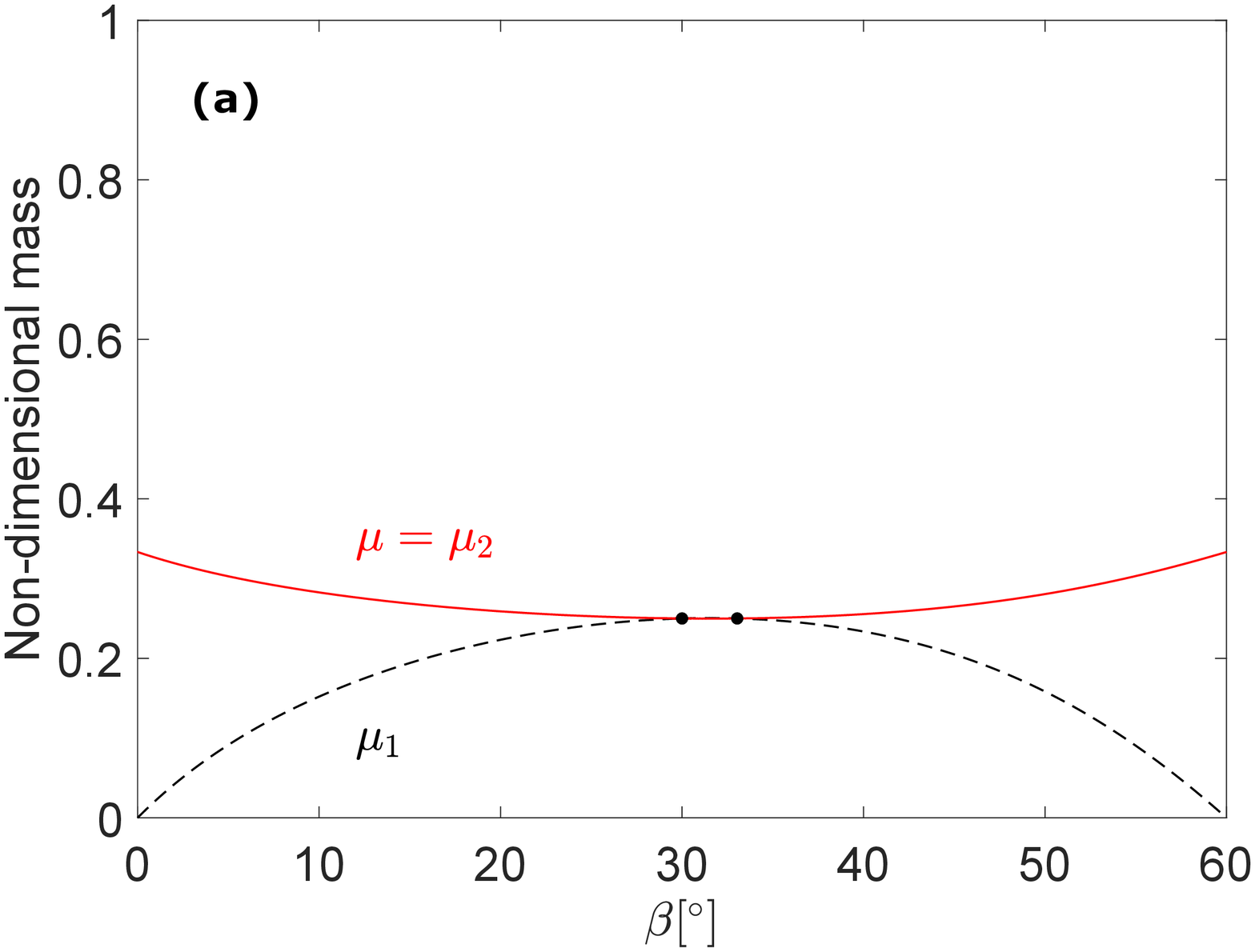}
\end{subfigure}
\quad
\begin{subfigure}[!h]{0.48\textwidth}
\includegraphics[width=\textwidth]{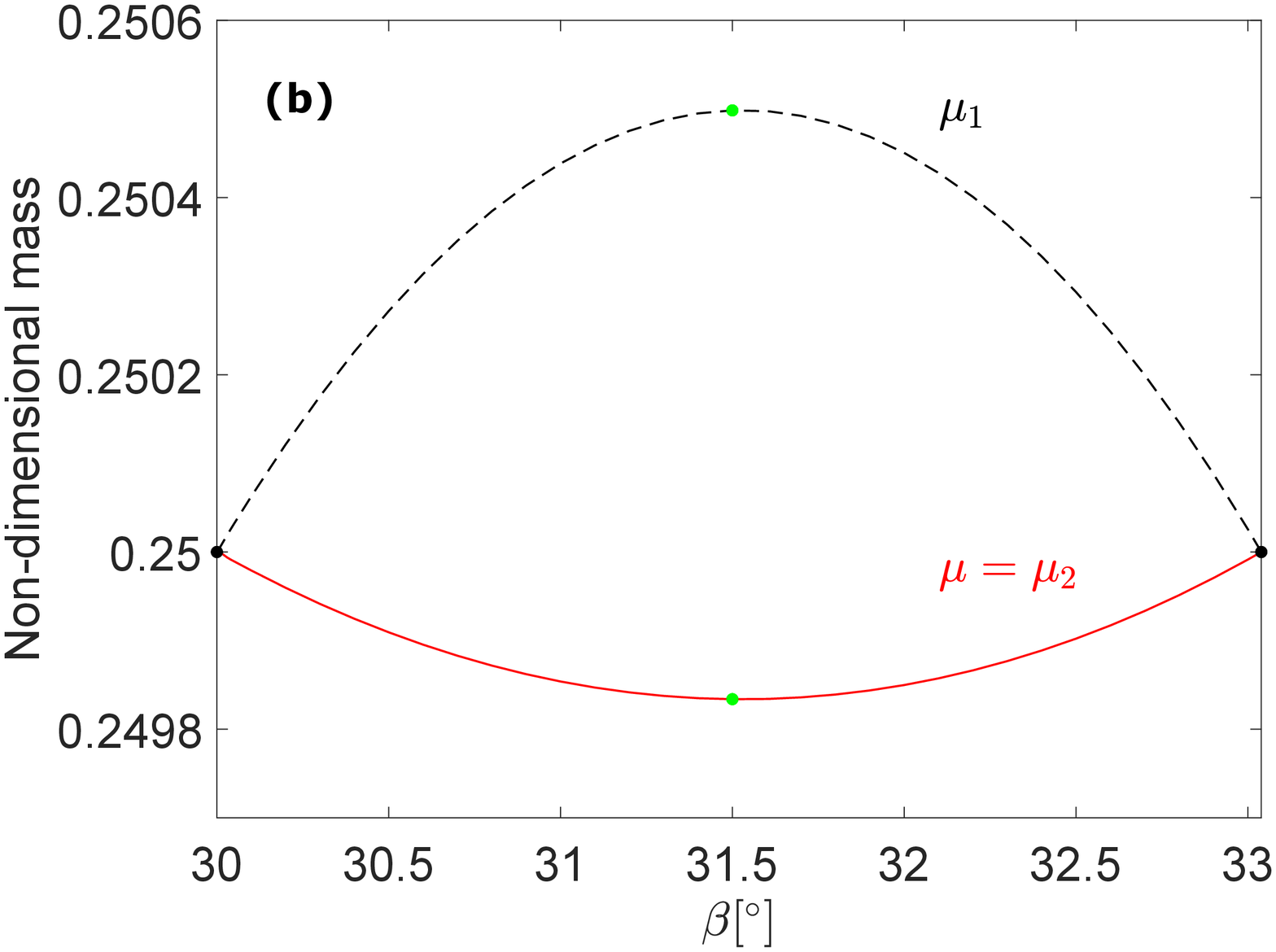}
\end{subfigure}
\caption{(\textbf{a}) Normalized masses in the function of the angle coordinate $ \beta $ (concave cases, $\mu = \mu_2 $). The~black dashed line refers to the residual mass $ \mu_1 $. The intersections at $ \beta = 30 \degree $ and at $ \beta = \text{33\arcdeg039} $ are emphasized by black dots and indicate the common mass $ \mu = \mu_1 = \mu_2 = 1/4 $. (\textbf{b}) A magnification of panel (\textbf{a}) in the region between the intersections. The ratio of the masses $ \mu_1 = 0.250499 $ and $ \mu = \mu_2 = 0.249834 $ at the green dots corresponds to the maximum $m^*$ of \cite{bernat2009}.}
\label{fig:beta_mass_cv2}
\end{figure}

{Figure} \ref{fig:beta_mass_cv2}a shows the $ \mu = \mu_2 $ case (red curve). As seen in Section~\ref{subsec:solutions_in_the_beta_alpha_plane_cv}, the mass of $ A $ has to be zero both at $ \beta = 0 \degree $ and $ 60 \degree $, and accordingly, the common mass $ \mu = \mu_2 $ takes the value $ 1/3 $ in these points (see {Figure} \ref{fig:beta_mass_cv2}a). Between these values of $\beta$, $ \mu_1 $ first monotonically increases, while $ \mu = \mu_2 $ slightly decreases. The first intersection of them occurs in the singular point $ S $ at $ \beta = 30 \degree $, where $ \mu = \mu_1 = \mu_2 = 1/4 $ is necessarily satisfied in the $ \mu = \mu_2 $ case (see Section~\ref{subsec:solutions_in_the_beta_alpha_plane_cv}). The second intersection at $\beta= \text{33\arcdeg039} $ also corresponds to the case of four equal masses (see also {Figure} \ref{fig:41_betaalphaplane_concaves_zoom}). In between the two intersections, the red curve $ \mu = \mu_2 $ reaches its global minimum and the curve $ \mu_1 $ its global maximum, as indicated by the green dots in {Figure} \ref{fig:beta_mass_cv2}b. The $ \alpha = \text{60\arcdeg593} $, $ \beta = \text{31\arcdeg529} $ coordinates corresponding to these extremal values of the masses can be determined as the minimum point of the function $ \beta \mapsto M(\beta) \equiv \mu_2(\beta)/\mu_1(\beta) $. For the details, see Appendix \ref{sec:appendix_c}. We note here that the ratio $ \mu_1/\mu_2 = 1.00266 $ at this minimum point corresponds to the maximal mass $ m^* $ of \cite{bernat2009}. After the second intersection, $ \mu_1 $ tends to the final value $ 0 $ as $ \mu = \mu_2 $ approaches $1/3$ and reaches it at the endpoint.

We also note that the singular point $ S $ (at $ \alpha = 60 \degree $, $ \beta = 30 \degree $) corresponds to the case when the bodies $ A $, $ E $, and $ E' $ form a Lagrangian equilateral triangle with equal masses (due to the condition $ 3 \mu_1 + \mu_2 = 1 $ for all values of $\mu_2$) and the body $B$ is situated in the center of the triangle. Since $\mu_2$ can take any value between $ 0 $ and $ 1 $ (accordingly, $ \mu = \mu_1 \in [0, 1/3] $), the masses indicated at $ \beta = 30 \degree $ in Figures \ref{fig:beta_mass_cv1} and \ref{fig:beta_mass_cv2} represent only two special cases among all the possible ones that can occur in the singular point $ S $.

\unskip

\section{Summary}
\label{sec:summary}
Central configurations bear great significance in the field of celestial mechanics. They represent the only subclass of the $ n $-body problem that promises analytic solutions. Yet, due to the rather complicated, multivariable equations describing them, only partial results are known for more than three bodies. With our work, we wish to have made progress in this challenging topic.

In the case of the four-body central configurations, a complete analytic solution was given recently~\cite{erdi2016} in the axisymmetric restriction of the problem, where two unequal-mass bodies (labeled by $ A $ and $ B $) define the axis of symmetry and the other two equal-mass bodies ($ E $ and $ E' $) are situated symmetrically with respect to this axis. Such configurations are called kite configurations. The positions of the bodies $ A $ and $ B $ are described by angle coordinates ($ \alpha $ and $ \beta $) with respect to those of $ E $ and $ E' $. According the the relative positions, we distinguish between convex and concave configurations.

Here, we applied the solutions of \cite{erdi2016} to systems containing three equal masses, and presented all such configurations with respect to the domain in the angle space of the kite central configurations. Taking advantage of the explicit nature of the results of \cite{erdi2016}, we gave a complete description of the problem by providing both the coordinates and the masses of the bodies.

We note here that the equilateral trapezoid central configurations are also axisymmetric but with two pairs of equal masses. Three equal masses are possible only when all four masses are equal and then the configuration turns into a square. Thus, our paper describes all axisymmetric four-body central configurations with three equal masses.

The three equal masses are achieved either by setting the mass of $ A $ to be equal to the mass of $ E $ and $ E' $ or by doing the same with $ B $. In both cases, we distinguished between the convex and concave configurations. In all cases, there are two families of solutions; corresponding to that either the mass-point $ A $ or the mass-point $ B $ has equal mass with the bodies $ E $ and $ E' $. In our solution, first we numerically computed those $ \alpha $, $ \beta $ pairs that satisfy the mass-equality conditions, then determined and studied the corresponding normalized masses. We repeat here the main characteristics of the solution.

In the convex $\mu=\mu_2$ case (when $B$, $E$, and $E'$ have equal masses), there exists an interval ($ \text{42\arcdeg211} < \alpha < \text{42\arcdeg640} $ and $ \text{23\arcdeg680} < \beta < \text{36\arcdeg125} $) in which for each value of $\alpha$ there are two values of $\beta$ satisfying the three-equal-mass condition. It means that for each position of $A$ there are two positions of $B$, such that $ B $, $ E $, and $ E' $ form equal-mass configurations.

In the concave $\mu=\mu_1$ case	(when $A$, $E$, and $E'$ have equal masses), there exists an interval ($ \text{56\arcdeg930} < \alpha < 60 \degree$ and $ \text{5\arcdeg678} < \beta < 30\degree$) too, in which for each value of $\alpha$ there are two values of $\beta$ satisfying the three-equal-mass condition. It means that for each position of $A$ there are two positions of $B$, such that $ A $, $ E $, and $ E' $ form equal-mass configurations.

In all the other convex and concave cases, the $ \alpha $, $ \beta $ pairs are unique, implying that for each position of $A$ there exists a unique position of $B$, such that one of them with $E$ and $E'$ provides the three-equal-mass part of the central configuration.

Our results are displayed in Figures \ref{fig:betaalphaplane_cx}, \ref{fig:betaalphaplane_cv}, \ref{fig:beta_mass_cx}, \ref{fig:beta_mass_cv1}, and \ref{fig:beta_mass_cv2}. The first two show the angle pairs of the three-equal-mass cases in the domain of the kite central configurations. The latter three present the corresponding masses. The four-equal-mass solutions are considered as special cases of the three equal masses. These occur at $ \alpha = \beta = 45 \degree $ in the convex case; and at $ \alpha = 60 \degree $, $ \beta = 30 \degree $; $ \alpha = \text{61\arcdeg177} $, $ \beta = \text{33\arcdeg039} $ in the concave cases.

Our solutions give a complete description of the three-equal-mass case of the axisymmetric four-body central configurations, relating them to the general case of such configurations. We have also shown the connection of our solutions with previous results.

\vspace{6pt}

\authorcontributions{Conceptualization, B.{\'E}.; formal analysis, E.K.; investigation, E.K.; writing---original draft preparation, E.K.; writing---review and editing, B.{\'E}.; supervision, B.{\'E}. All authors have read and agreed to the published version of the manuscript.}

\funding{This research was supported by the ÚNKP-19-3 New National Excellence Program of the Ministry for Innovation and Technology.}

\acknowledgments{The authors thank the reviewers for their supportive comments.}

\conflictsofinterest{The authors declare no conflict of interest.}

\appendixtitles{yes} 
\appendix
\section{The Coefficients  \emph{a}\textsubscript{0}, \emph{a}\textsubscript{1}, \emph{b}\textsubscript{0}, \emph{b}\textsubscript{1}}
\label{sec:appendix_a}
In the convex case:
\begin{eqnarray*}
a_0 &=& \left(\cos^3\alpha-\frac{1}{8}\right)\tan\alpha, \label{Cxa0} \\
a_1 &=& \frac{1}{\left(\tan\alpha+\tan\beta\right)^2}+\left(\frac{1}{8}-\cos^3\alpha-\cos^3\beta\right)\tan\beta-\frac{1}{8}\tan\alpha, \label{Cxa1} \\
b_0 &=& \left(\cos^3\beta-\frac{1}{8}\right)\tan\beta, \label{Cxb0} \\
b_1 &=& \frac{1}{\left(\tan\alpha+\tan\beta\right)^2}+\left(\frac{1}{8} -\cos^3\alpha-\cos^3\beta\right)\tan\alpha-\frac{1}{8}\tan\beta. \label{Cxb1}
\end{eqnarray*}
\par
In the concave cases:
\begin{eqnarray*}
a_0 &=& \left(\cos^3\alpha-\frac{1}{8}\right)\tan\alpha, \label{Cca0} \\
a_1 &=& \frac{1}{\left(\tan\alpha-\tan\beta\right)^2}-\left(\frac{1}{8}-\cos^3\alpha-\cos^3\beta\right)\tan\beta-\frac{1}{8}\tan\alpha, \label{Cca1} \\
b_0 &=& -\left(\cos^3\beta-\frac{1}{8}\right)\tan\beta, \label{Ccb0} \\
b_1 &=& \frac{1}{\left(\tan\alpha-\tan\beta\right)^2}+\left(\frac{1}{8} -\cos^3\alpha-\cos^3\beta\right)\tan\alpha+\frac{1}{8}\tan\beta. \label{Ccb1}
\end{eqnarray*}

\section{Properties of the Functions \emph{f\textsubscript{i}}  (\emph{i} = 1, \dotso , 4)}
\label{sec:appendix_b}
\vspace{-6pt}
\subsection{Function $ f_1 $ (Concave Cases, $ \mu = \mu_1 $)}
\label{subsec:appendix_b1}
Function $ f_1 $: $ B_1 \rightarrow A_1 $ is defined implicitly by Equation \eqref{eq:muequalsmu1convexfinal} and represented by the blue curve in {Figure} \ref{fig:betaalphaplane_cx}. Its domain $ B_1 \cong [45 \degree, 53 \degree] $ and codomain $ A_1 = [45 \degree, 60 \degree] $ refer to the intervals of the angle coordinates $ \beta $ and $ \alpha $, respectively.

We claim that $ f_1 $ is injective. For this, it is enough to prove that $ f_1 $ is strictly monotone in $ B_1 $. One must calculate, therefore, the derivative $ g_1 \equiv \mathrm{d}f_1/\mathrm{d}\beta $ and show that it is definite.

By totally differentiating Equation \eqref{eq:muequalsmu1convexfinal} with respect to $ \beta $, it yields, after some arrangements, that
\begin{equation}
\label{eq:g1}
g_1(\beta) = \frac{N_1(\beta)}{D_1(\beta)}, \nonumber
\end{equation}
where the numerator $ N_1 $ and denominator $ D_1 $ take the forms
\begin{eqnarray*}
\label{eq:g1}
N_1(\beta) &=& 2\cos\beta\left(3\sin^2\beta-1\right) + \frac{1}{\cos^2\beta}\left(\frac{1}{4}+\cos^3\alpha+\frac{2}{\left(\tan\alpha+\tan\beta\right)^3}\right),  \\
D_1(\beta) &=& \cos\alpha\left(3\sin^2\alpha-1\right) + 3\sin\alpha\cos^2\alpha\tan\beta - \frac{2}{\left(\tan\alpha+\tan\beta\right)^3\cos^2\alpha},
\end{eqnarray*}
where all $ \alpha \equiv \alpha(\beta) $. We show that $ g_1 $ is positive for all $ \alpha \in A_1 $, $ \beta \in B_1 $.

In the numerator $ N_1 $, it is easily seen that the 2nd term is positive. As for the 1st term, \mbox{$ 2\cos\beta > 0 $ and}
\begin{equation}
3\sin^2\beta-1 \geq \min_{\beta\in B_1}\left\{3\sin^2\beta-1\right\} = 3\sin^2 45 \degree-1 = \frac{1}{2} > 0. \nonumber
\end{equation}

Thus, $ N_1 > 0 $.

In the denominator $ D_1 $, the positive sign of the 1st term can be seen similarly to that of the 1st term in $ N_1 $, but using $ \alpha $ instead of $ \beta $. In the sum of the 2nd and 3rd terms:
\begin{multline}
3\sin\alpha\cos^2\alpha\tan\beta - \frac{2}{\left(\tan\alpha+\tan\beta\right)^3\cos^2\alpha} \geq \min_{\beta\in B_1}\left\{3\sin\alpha\cos^2\alpha\tan\beta - \frac{2}{\left(\tan\alpha+\tan\beta\right)^3\cos^2\alpha}\right\}  \\
= 3\sin\alpha\cos^2\alpha\tan45\degree - \frac{2}{\left(\tan\alpha+\tan45\degree\right)^3\cos^2\alpha} = 3\sin\alpha\cos^2\alpha - \frac{2}{\left(\tan\alpha+1\right)^3\cos^2\alpha}, \nonumber
\end{multline}
where the last expression is a strictly, monotonically decreasing one-variable function of $ \alpha $ (since its derivative is negative for all $ \alpha \in A_1 $). Hence, it is minimized by substituting the maximal $ \alpha = 60 \degree $, which gives $ \sim 0.257 > 0 $. Thus, $ D_1 > 0 $ too, and this proves the above claim as well.

\subsection{Function $ f_2 $ (Concave Cases, $ \mu = \mu_2 $)}
\label{subsec:appendix_b2}
Function $ f_2 $: $ B_2 \rightarrow A_2 $ is defined implicitly by Equation \eqref{eq:muequalsmu2convexfinal} and represented by the red curve in {Figure} \ref{fig:betaalphaplane_cx}. Its domain $ B_2 \cong [23 \degree, 45 \degree] $ and codomain $ A_2 \cong [42 \degree, 45 \degree] $ refer to the intervals of the angle coordinates $ \beta $ and $ \alpha $, respectively.

We claim that $ f_2 $ is noninjective. For this, it is enough to prove that $ f_2 $ has a local extremum in $ B_2 $. One must show, therefore, that the first derivative $ g_2 \equiv \mathrm{d}f_2/\mathrm{d}\beta $ has a root in some point of $ B_2 $ and that the second derivative $ h_2 \equiv \mathrm{d}g_2/\mathrm{d}\beta \equiv \mathrm{d}^2f_2/\mathrm{d}\beta^2 $ is either positive or negative in this point.

By totally differentiating Equation \eqref{eq:muequalsmu2convexfinal} with respect to $ \beta $, it yields, after some arrangements, that
\begin{equation}
\label{eq:g1}
g_2(\beta) = \frac{N_2(\beta)}{D_2(\beta)}, \nonumber
\end{equation}
where the numerator $ N_2 $ and denominator $ D_2 $ take the forms
\begin{eqnarray*}
\label{eq:g1}
N_2(\beta) &=& \cos\beta\left(3\sin^2\beta-1\right) + 3\sin\beta\cos^2\beta\tan\alpha - \frac{2}{\left(\tan\alpha+\tan\beta\right)^3\cos^2\beta},  \\
D_2(\beta) &=& 2\cos\alpha\left(3\sin^2\alpha-1\right) + \frac{1}{\cos^2\alpha}\left(\frac{1}{4}+\cos^3\beta+\frac{2}{\left(\tan\alpha+\tan\beta\right)^3}\right),
\end{eqnarray*}
where all $ \alpha \equiv \alpha(\beta) $.

First, we show that the denominator $ D_2 $ is nonzero (positive) for all $ \alpha \in A_2 $, $ \beta \in B_2 $. The positive sign of the 2nd term is easily seen. As for the 1st term, $ 2\cos\alpha > 0 $ and
\begin{equation}
3\sin^2\alpha-1 \geq \min_{\alpha\in A_2}\left\{3\sin^2\alpha-1\right\} = 3\sin^2 42 \degree-1 \cong 0.343 > 0. \nonumber
\end{equation}

Thus, $ D_2 \neq 0 $.

Next, we proceed to determine the root of the numerator $ N_2 $. Since this problem is analytically unfeasible, we used \emph{MATLAB} to find the root, which satisfies Equation \eqref{eq:muequalsmu2convexfinal} as well. The result is $ \alpha = \text{42\arcdeg211} $, $ \beta = \text{30\arcdeg154} $. In this point, the second derivative $ h_2 $ is positive, implying the type of extremum to be a local minimum.

\subsection{Function $ f_3 $ (Concave Cases, $ \mu = \mu_1 $)}
\label{subsec:appendix_b3}
Function $ f_3 $: $ B_3 \rightarrow A_3 $ is defined implicitly by Equation \eqref{eq:muequalsmu1concavefinal} and represented by the blue curve in {Figure} \ref{fig:betaalphaplane_cv}. Its domain $ B_3 \cong [5 \degree, 49 \degree] $ and codomain $ A_3 \cong [56 \degree, 70 \degree] $ refer to the intervals of the angle coordinates $ \beta $ and $ \alpha $, respectively.

We claim that $ f_3 $ is noninjective. For this, it is enough to prove that $ f_3 $ has a local extremum in $ B_3 $. One must show, therefore, that the first derivative $ g_3 \equiv \mathrm{d}f_3/\mathrm{d}\beta $ has a root in some point of $ B_3 $ and that the second derivative $ h_3 \equiv \mathrm{d}g_3/\mathrm{d}\beta \equiv \mathrm{d}^2f_3/\mathrm{d}\beta^2 $ is either positive or negative in this point.

By totally differentiating Equation \eqref{eq:muequalsmu1concavefinal} with respect to $ \beta $, it yields, after some arrangements, that
\begin{equation}
\label{eq:g3}
g_3(\beta) = \frac{N_3(\beta)}{D_3(\beta)}, \nonumber
\end{equation}
where the numerator $ N_3 $ and denominator $ D_3 $ take the forms
\begin{eqnarray*}
\label{eq:g1}
N_3(\beta) &=& 2\cos\beta\left(3\sin^2\beta-1\right) + \frac{1}{\cos^2\beta}\left(\frac{1}{4}+\cos^3\alpha+\frac{2}{\left(\tan\alpha-\tan\beta\right)^3}\right),  \\
D_3(\beta) &=& 3\sin\alpha\cos^2\alpha\tan\beta + \cos\alpha - 3\sin^2\alpha\cos\alpha + \frac{2}{\left(\tan\alpha-\tan\beta\right)^3\cos^2\alpha},
\end{eqnarray*}
where all $ \alpha \equiv \alpha(\beta) $.

First, we show that the denominator $ D_3 $ is nonzero (positive) for all $ \alpha \in A_3 $, $ \beta \in B_3 $:
\begin{equation}
D_3 \geq \min_{\beta \in B_3} \left\{D_3\right\} = 3\sin\alpha\cos^2\alpha\tan5\degree + \cos\alpha - 3\sin^2\alpha\cos\alpha + \frac{2}{\left(\tan\alpha-\tan5\degree\right)^3\cos^2\alpha}, \nonumber
\end{equation}
which is a strictly, monotonically decreasing one-variable function of $ \alpha $ (since its derivative is negative for all $ \alpha \in A_3 $). Hence, it is minimized by substituting the maximal $ \alpha = 70 \degree $, which gives $ \sim0.373 > 0 $. Thus, $ D_3 \neq 0 $.

Next, we proceed to determine the root of the numerator $ N_3 $. Since this problem is analytically unfeasible, we used \emph{MATLAB} to find the root, which satisfies Equation \eqref{eq:muequalsmu1concavefinal} as well. The result is $ \alpha = \text{56\arcdeg930} $, $ \beta = \text{15\arcdeg414} $. In this point, the second derivative $ h_3 $ is positive, implying the type of extremum to be a local minimum.

We observe that this minimum of $ f_3 $ is in the first concave region ($ \beta < 30 \degree $, $ \alpha < 60 \degree $). We claim that in the second concave region, $ f_3 $ is injective (strictly, monotonically increasing). The aim, therefore, is to show that the derivative $ g_3 $ is positive for all $ \beta \in [30 \degree, 49 \degree] $, $ \alpha \in [60 \degree, 70 \degree] $. It is sufficient for this if $ N_3 > 0 $, since the positive sign of the denominator $ D_3 $ applies also in the second concave region.

We start by minimizing $ N_3 $ with respect to $ \alpha $ by substituting $ \alpha = 70 \degree $. The resulting expression is a strictly monotonically increasing one-variable function of $ \beta $ (since its derivative is positive for all $ \beta \in [30 \degree, 49 \degree] $). Hence, we can minimize again, now with respect to $ \beta $, by substituting $ \beta = 30 \degree $. This gives $ \sim 0.215 > 0 $. Thus, in the second concave region $ N_3 > 0 $, which proves the claim.

\subsection{Function $ f_4 $ (Concave Cases, $ \mu = \mu_2 $)}
\label{subsec:appendix_b4}
Function $ f_4 $: $ B_4 \rightarrow A_4 $ is defined implicitly by Equation \eqref{eq:muequalsmu2concavefinal2} and represented by the red curve in {Figure} \ref{fig:betaalphaplane_cv}. Its domain $ B_4 = [0 \degree, 60 \degree] $ and codomain $ A_4 \cong [48 \degree, 72 \degree] $ refer to the intervals of the angle coordinates $ \beta $ and $ \alpha $, respectively.

We claim that $ f_4 $ is injective. For this, it is enough to prove that $ f_4 $ is strictly monotone in $ B_4 $. One must calculate, therefore, the derivative $ g_4 \equiv \mathrm{d}f_4/\mathrm{d}\beta $ and show that it is definite.

By totally differentiating Equation \eqref{eq:muequalsmu2concavefinal2} with respect to $ \beta $, it yields, after some arrangements, that
\begin{equation}
\label{eq:g1}
g_4(\beta) = \frac{N_4(\beta)}{D_4(\beta)}, \nonumber
\end{equation}
where the numerator $ N_4 $ and denominator $ D_4 $ take the forms
\begin{eqnarray*}
N_4(\beta) &=& 3\sin\beta\cos\beta\left(\cos\beta\tan\alpha-\sin\beta\right) + \cos\beta + \frac{2}{\left(\tan\alpha-\tan\beta\right)^3\cos^2\beta},  \\
D_4(\beta) &=& 2\cos\alpha\left(3\sin^2\alpha-1\right) + \frac{1}{\cos^2\alpha}\left(\frac{1}{4}+\cos^3\beta+\frac{2}{\left(\tan\alpha-\tan\beta\right)^3}\right),
\end{eqnarray*}
where all $ \alpha \equiv \alpha(\beta) $. We show that $ g_4 $ is positive for all $ \alpha \in A_4 $, $ \beta \in B_4 $, $ \alpha > \beta $ (a constraint in the concave cases).

In the numerator $ N_4 $, it is easily seen that the 2nd and 3rd terms are positive. As for the 1st term, $ 3\sin\beta\cos\beta > 0 $ and $ \cos\beta\tan\alpha - \sin\beta = \cos\beta\left(\tan\alpha-\tan\beta\right) > 0 $ (since $ \alpha > \beta $). Thus, $ N_4 > 0 $.

In the denominator $ D_4 $, the positive sign of the 2nd term is trivial, while in the 1st term \scalebox{.95}[1.0]{$ 2\cos\alpha > 0 $} and
\begin{equation}
3\sin^2\alpha - 1 \geq \min_{\alpha \in A_4}\left\{3\sin^2\alpha - 1\right\} = 3\sin^2 48 \degree - 1 \cong 0.657 > 0. \nonumber
\end{equation}

Thus, $ D_4 > 0 $ too, and this proves the above claim as well.

\section{The Minimum Point of Function \emph{M} }
\label{sec:appendix_c}
Function $ M $ is defined in the concave, $ \mu = \mu_2 $ case as the ratio of the functions $ \mu_2 $ and $ \mu_1 $:
\begin{equation}
\label{eq:M}
M(\beta) \equiv \frac{\mu_2(\beta)}{\mu_1(\beta)} = -\frac{(a_0-b_0-a_1)a_0}{(a_0-b_0+b_1)b_0}, \nonumber
\end{equation}
where $ (a_0-b_0+b_1)b_0 \neq 0 $ in the domain of $ M $, i.e., for $ \beta \in [30\degree, \text{33\arcdeg093}] $. Substituting the coefficients $ a_0 $, $ a_1 $, $ b_0 $, $ b_1 $ from Appendix \ref{sec:appendix_a} and using also Equation \eqref{eq:muequalsmu2concavefinal2} as the condition for the relation of $ \beta $ and $ \alpha(\beta) $, one obtains after some arrangements that
\begin{equation}
\label{eq:Mfinal}
M(\beta) = \frac{\cos^3\alpha\left(3\tan\alpha-\tan\beta\right) - \cos^3\beta\left(\tan\alpha-\tan\beta\right) - \frac{1}{4}\tan\alpha}{-2\cos^3\beta\tan\beta + \frac{1}{4}\tan\beta}, \nonumber
\end{equation}
where once again, the denominator is nonzero in the domain of $ M $ and all $ \alpha \equiv \alpha(\beta) $.

We claim that $ M $ has a minimum point. To prove this, one must calculate the derivative $ \mathrm{d}M/\mathrm{d}\beta $, which contains $ \beta $, $ \alpha(\beta) $, and $ \mathrm{d}(\alpha(\beta))/\mathrm{d}\beta $ too. With the observation, however, that $ \mathrm{d}(\alpha(\beta))/\mathrm{d}\beta = g_4(\beta) $ (see Appendix \ref{subsec:appendix_b4}), we can eliminate the last dependence and arrive to the form
\begin{equation}
\label{eq:Mder}
\frac{\mathrm{d}M(\beta)}{\mathrm{d}\beta} = \frac{N_{51}(\beta)N_{52}(\beta) - N_{53}(\beta)N_{54}(\beta)}{D_5(\beta)}, \nonumber
\end{equation}
where
\begin{eqnarray*}
N_{51}(\beta) &=& g_4(\beta)\left(3\cos\alpha - 3\sin\alpha\cos^2\alpha\left(3\tan\alpha-\tan\beta\right) - \frac{1}{\cos^2\alpha}\left(\cos^3\beta + 1\right)\right) \\
&& + 3\sin\beta\cos^2\beta\left(\tan\alpha - \tan\beta\right) - \frac{\cos^3\alpha}{\cos^2\beta} + \cos\beta, \\
N_{52}(\beta) &=& -2\cos^3\beta\tan\beta + \frac{1}{4}\tan\beta, \\
N_{53}(\beta) &=& \cos^3\alpha\left(3\tan\alpha-\tan\beta\right) - \cos^3\beta\left(\tan\alpha-\tan\beta\right) - \frac{1}{4}\tan\alpha, \\
N_{54}(\beta) &=& 2\cos\beta\left(3\sin^2\beta - 1\right) + \frac{1}{4\cos^2\beta}, \\
D_{5}(\beta) &=& \tan^2\beta \left(4\cos^6\beta - \cos^3\beta + \frac{1}{16}\right),
\end{eqnarray*}
where $ D_5 \neq 0 $ in the domain of $ M $ and all $ \alpha \equiv \alpha(\beta) $. Finding the root of $ \mathrm{d}M/\mathrm{d}\beta $ is analytically unfeasible, but the numerical approach gives $ \alpha = \text{60\arcdeg593} $, $ \beta = \text{31\arcdeg529} $ as the unique solution that satisfies Equation \eqref{eq:muequalsmu2concavefinal2} as well. The positive sign of the second derivative $ \mathrm{d}^2M/\mathrm{d}\beta^2 $ in this point provides the sufficient condition for the initial claim.

\reftitle{References}





\end{document}